\documentclass[twoside]{amsart}
\usepackage{latexsym}
\usepackage{amssymb,amsmath,amsopn}
\usepackage[dvips]{graphicx}   
\usepackage{color,epsfig}      
\usepackage{subcaption}
\usepackage{amsbsy}
\usepackage{multirow}
\usepackage{epstopdf}
\usepackage{rotating}
\usepackage{adjustbox}
\usepackage{lineno}

               {\begin{list}{}{\leftmargin#1\rightmargin#2}\item{}}%
               {\end{list}}

\def\xR{\mathbb{R}}

\def\diag{\textnormal{diag}}
\def\xe{\mathbf{e}}
\def\xg{\mathbf{g}}
\def\xi{\mathbf{i}}

\def\xk{\mathbf{k}}
\def\xu{\mathbf{u}}
\def\xA{\mathbf{A}}
\def\xB{\mathbf{B}}
\def\xD{\mathbf{D}}

\def\xI{\mathbf{I}}
\def\xH{\mathbf{H}}

\def\xM{\mathbf{M}}
\def\xQ{\mathbf{Q}}

\def\xV{\mathbf{V}}
\def\xW{\mathbf{W}}
\def\xx{\mathbf{x}}

\def\xq{\mathbf{q}}

\def\xs{\mathbf{s}}
\def\xn{\mathbf{n}}
\def\xt{\mathbf{t}}
\def\xu{\mathbf{u}}

\def\xx{\mathbf{x}}

\def\xDelta{\boldsymbol{\Delta}}
\def\xdelta{\boldsymbol{\delta}}

\def\xsigma{\boldsymbol{\sigma}}

\begin{document}
\title[Integrated magnetic and mesotectonic data evaluation]{Reconstruction of early phase deformations by integrated magnetic and mesotectonic data evaluation}
\author[A. \'A. Sipos, E. M\'arton \and L. Fodor]{Andr\'as \'Arp\'ad Sipos, Em\H{o} M\'arton \and L\'aszl\'o Fodor}

\address{Andr\'as \'A. Sipos, MTA-BME Morphodynamics Research Group and Dept. of Mechanics, Materials and Structures, Budapest University of Technology and Economics, Budapest, Hungary}
\address{Em\H{o} M\'arton, Mining and Geological Survey of Hungary, Paleomagnetic Laboratory, Budapest, Hungary}
\address{L\'aszl\'o Fodor, MTA-ELTE Geological, Geophysical and Space Science Research Group, Budapest, Hungary}
\email{siposa@eik.bme.hu}

\keywords{Magnetic fabrics, rotational anisotropy, stochastic stress inversion, reconstruction of weak deformations}

\begin{abstract}
Markers of brittle faulting are widely used for recovering past deformation phases. Rocks often have oriented magnetic fabrics, which can be interpreted as connected to ductile deformation before cementation of the sediment. This paper reports a novel statistical procedure for simultaneous evaluation of AMS (Anisotropy of Magnetic Susceptibility) and fault-slip data.The new method analyzes the AMS data, without linearization techniques, so that weak AMS lineation and rotational AMS can be assessed that are beyond the scope of classical methods. This idea is extended to the evaluation of fault-slip data. While the traditional assumptions of stress inversion are not rejected, the method recovers the stress field via statistical hypothesis testing. In addition it provides statistical information needed for the combined evaluation of the AMS and the mesotectonic (0.1 to 10m) data. In the combined evaluation a statistical test is carried out that helps to decide if the AMS lineation and the mesotectonic markers (in case of repeated deformation of the oldest set of markers) were formed in the same or different deformation phases. If this condition is met, the combined evaluation can improve the precision of the reconstruction. When the two data sets do not have a common solution for the direction of the extension, the deformational origin of the AMS is questionable. In this case the orientation of the stress field responsible for the AMS lineation might be different from that which caused the brittle deformation. Although most of the examples demonstrate the reconstruction of weak deformations in sediments, the new method is readily applicable to investigate the ductile-brittle transition of any rock formation as long as AMS and fault-slip data are available.
\end{abstract}
\maketitle

\section{Introduction}
\label{Intro}

Reconstruction of the former orientations of past deformations of geological units is one of the key questions in the geosciences. In several cases the small amount of overall deformation is reflected in only a few, weak markers making historical analysis difficult, often impossible. The ductile to brittle sequence of deformation styles is widely presumed during the deformation history for most rocks (lithifying sediments, cooling magmatic and some metamorphic rocks). If the basic cause of the deformation -- namely stress -- prevails beyond the early (ductile) phase of deformation, then it might lead to brittle fracture (faults, joints, deformation bands) in the rock unit \cite{Tal}. Our work aims to approach this transition, in particular cases, when it takes place in a predominantly steady stress field. An integrated method that facilitates two, frequently available indicators, and exploits relatively low range of deformation, might shed light on the transitional field of the ductile and brittle deformation styles. 

Both magnetic fabric (AMS, Anisotropy of Magnetic Susceptibility) and mesotectonic markers are widely used for reconstructing past deformation phases (following \cite{tuwe} and \cite{han}, the mesotectonic scale refers to the range between 0.1m and 10m). Although later deformation phases may occur, this transition phase is unique as it is the only one that is reflected by both quasi-simultaneous magnetic and mesotectonic markers. In the terms of continuum mechanics, we thus consider the first increment of the strain.

In both AMS and fault-slip methods there are doubts about whether the directions of the stress field is reflected more precisely in AMS or in brittle deformations. Some studies \cite[e.g.][]{hae}) point out that AMS is an unreliable predictor of not only stress, but even strain. Others simulate well-defined multiphysical models and demonstrate the highly nonlinear dependence of the susceptibility tensor on the finite strain during successive events of deformations \cite{jez}. Undoubtedly, such observations and models must be valid for the general situation in which any material under any specific deformation is distorted to an arbitrary extent. However, in the case of weak deformation of homogeneous sediments, a correlation has been demonstrated between stress (reflected by brittle deformation markers) and AMS data (\cite{bo2,ci,fe} and references therein). These studies, in principle, state that the formation of the AMS fabric takes place during the early, unconsolidated stage. 

The intuitive physical picture outlined above relies on the following assumptions:
\begin{itemize}
	\item the AMS reflects the weak deformation of the early, ductile phase, prior to advanced lithification;
	\item the cause of the deformation lasted sufficiently to produce brittle markers;
	\item in sedimentary rocks, the deformation happened while the layers were horizontal.
\end{itemize}

Unfortunately, even if the above criteria are met, statistical analysis is difficult because we are dealing with weak deformations and both AMS and mesotectonic markers are sparse. So the available data tend to be noisy, making statistical treatment of such data-sets uncertain. In the case of tensor quantities some linearization technique can usually be applied to statistically evaluate eigendirections of the tensor \cite[e.g.][]{Cai}). If the eigendirections are considered as independent vectors, then procedures developed for vectors can be used, such as Fisher statistics over the sphere \cite{fi}, or its modified version by \cite{bing} or \cite{hn}. Random sampling with replacement known as ``bootstrapping" might help to overcome the difficulties of small sample size or unknown distribution type \cite{Tx2,Tx1}. Several authors point out that these approaches completely neglect the tensor nature of the observed quantities \cite{tx}. Methods, which aim to keep consistency with the underlying physics strongly rely on linearization techniques \cite{he,je1}, but as pointed out in \cite{he}, the error due to the linearization (i.e. neglecting higher order terms in the Taylor series of a tensor) can be quite large, hence the approximation of the confidence intervals might be poor. It is not difficult to see that two, sufficiently close eigenvalues of the tensor (which situation is referred to as \emph{rotational anisotropy} throughout the paper) lead to the underestimation of the confidence regions by any method built on linearization. 

In this paper a statistical framework for tensor quantities is presented that -- apart from a mild assumption about normal distribution of the input data -- is free from other \emph{a-priori} assumptions (i.e. it is able to handle data-sets represented by closely rotationally anisotropic tensors), and the accuracy of the computed confidence intervals does not depend on intrinsic characteristics of the outcome (such as the degree of AMS lineation). 

Our approach is readily applicable for AMS data sets and can be extended to the stress inversion applied in mesotectonics. The idea of using both sources simultaneously in reconstructing the orientation of past stress field is common practice and relies heavily on visual comparison of stereograms and hence biased by human intuition. The new method of combined statistical evaluation of the AMS and mesotectonic data can be applied to several  kinds of geological objects. It can be used to study the ductile to brittle transition and investigate the steadiness of the stress field. However, it is particularly powerful when the maximum and intermediate axes of the AMS ellipsoid are of similar length, as in moderately deformed samples of soft and fine grained sediments, and where the availability of the mesotectonic data is limited.

Although this paper is devoted to the statistical procedure itself, the methods to obtain AMS and mesotectonic data will also be discussed briefly and the applicability of the method will be demonstrated using field examples from the Pannonian basin, Central Europe.

\subsection{AMS measurements and the interpretation of the results in terms of deformation}
The AMS ellipsoid is determined on oriented field samples. The magnetic susceptibility tensor for each sample is measured on different instruments \cite{st}. During the measurement the sample is placed in a magnetic field ($\xH$) and its magnetization ($\xM$) is determined for several spatial orientations. The magnetic susceptibility tensor describes the linear transformation between the vectors $\xH$ and $\xM$ via $\xM=\xk\xH$. It can be represented by a $3\times 3$, symmetric, real valued matrix, 
\begin{eqnarray}
	\mathbf{k}=
	\begin{bmatrix}
	k_{11} & k_{12} & k_{13} \\
	k_{12} & k_{22} & k_{23} \\
	k_{13} & k_{23} & k_{33} 
	\end{bmatrix}.
\end{eqnarray}

Several devices and testing procedures are available to carry out the measurements, details for which are provided by \cite{je2} and \cite{st}, and references therein. The AMS ellipsoid characterizes the magnetic fabric of a rock. It is considered as \emph{primary} in a sediment formed during deposition and, in igneous rocks, during cooling in the absence of  external forces.  In sediments, the AMS ellipsoid is oblate, the orientation of the maximal principal axis (denoted as K1) extends over a wide range of azimuths, sometimes even in a single layer, but always throughout a stratigraphic sequence, due to the temporal changes of the flow direction within the sedimentary basin. In some cases a general trend can be observed that is maintained throughout a stratigraphic sequence, especially  in the  fine grained clastics (mudstones). This trend can be attributed to \emph{weak tectonic deformation} \cite{mat,ci,ma,ma1,ma2}, especially when K3 is close to the bedding pole, i.e. the magnetic foliation is subparallel with the bedding plane. The deformation leaving a magnetic imprint in these sediments is primary, the first one after the deposition. Overprinting of this early AMS fabric by subsequent tectonic phases is unlikely, as the magnetic fabric of the sediment more readily reflects strain while the sediment is relatively soft, i.e., able to undergo continuous (ductile) deformation and did not go through cementation process \cite{bo1}. The magnetic fabrics of igneous (lava) rocks can be affected by strain while they are not yet completely cooled \cite{ma,les}. Afterward which their fabrics are difficult to modify \cite{Ta}.

\subsection{Methodology of fault-slip analysis}
Field measurements generally comprise the measurement of strike and dip data of striated fault planes, joints, deformation bands or other types of brittle elements. Fault kinematics can be determined using divers criteria described in several papers \cite{ang4,han,pet}. Starting from fault-slip data several algorithms were elaborated for calculation of the $\xsigma$ stress tensor \cite{ang,ang2,za1,za2}. In most cases only the reduced stress tensor is determined incorporating the orientation of stress axes and their ratio, but not their absolute value \cite{carey}.

In the case of multiple faulting phases, a combination of automatic \cite{ang6} and manual separation, or their combination, can be used to separate faults into phases. Some of the data in this paper were analyzed in a combined way \cite{sib,fo}. The tilt test is useful and important for sedimentary rocks in order to establish the relative chronology between faulting and tilting around a  horizontal axis. For a conjugate set of faults, that underwent tilting, the symmetry plane of faults and also the stress axes deviate from vertical and horizontal; thus backtilting of faults to their horizontal bed position would reconstruct the original position of the stress axes at the time of faulting. Although the tilting itself and the faulting could belong to the same deformation phase, these successive events could be coaxial. Early faulting while in a horizontal bed position and tilting could equally be separated in time and characterized by different stress/strain axes.

\begin{figure}
\includegraphics[width=0.50\textwidth]{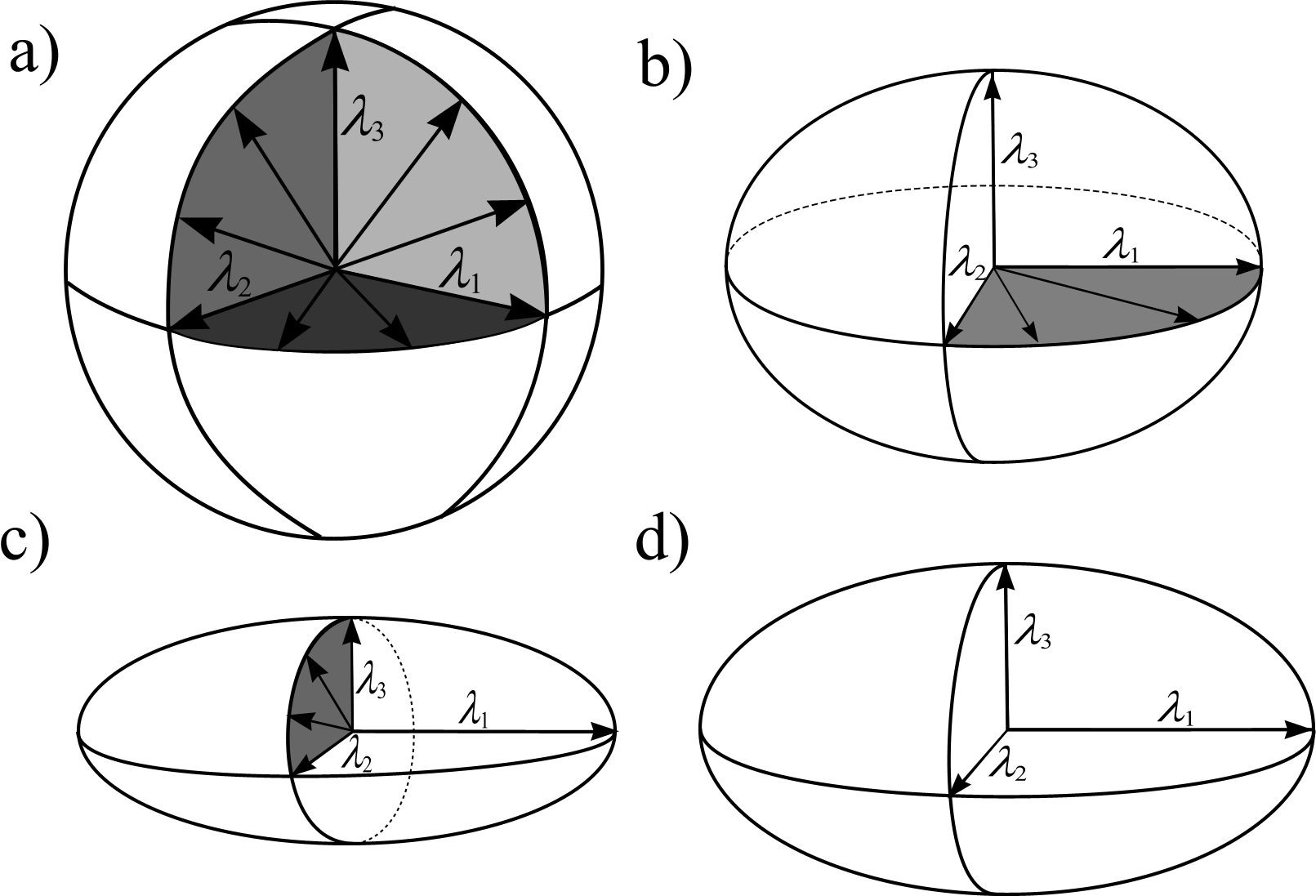}
\caption{Ellipsoids associated with $3\times 3$, positive definite tensors: a) \emph{sphere} ($\lambda_1=\lambda_2=\lambda_3$) isotropic; b) \emph{oblate spheroid} ($\lambda_1=\lambda_2>\lambda_3$) rotationally anisotropic; c) \emph{prolate spheroid} ($\lambda_1>\lambda_2=\lambda_3$) rotationally anisotropic; d) \emph{ellipsoid} ($\lambda_1>\lambda_2>\lambda_3$) anisotropic.}
\label{fig:00}
\end{figure}

\subsection{Assumptions}
We aim to treat cases in which the first deformation phase induced the magnetic fabric (at grain scale) as well as causing brittle fracturing (at meter scale) in a practically horizontal position. We restrict ourselves to the following assumptions. Let $\xk_{i,0}$ denote the specimen magnetic susceptibility tensor. $\xk_{i,0}$ is determined with a negligible error (i.e. the maximal semi-major axis of the confidence-ellipse of the principal directions, $E_{12}<15^\circ$) and its elements are positive reals, hence the tensor can be associated with an ellipsoid (Figure \ref{fig:00}). A locality is represented by $N$ pieces of oriented samples. Even within one locality the volumes of the ellipsoids may differ. Since we aim to analyze the eigendirections of the resultant tensor, normalization of all measured tensor is desirable. To be consistent with normal practice, normalization is carried out by the first scalar invariant $I_1$ of $\xk$, namely
\begin{eqnarray}
\label{eq:normalization}
\xk_i=\frac{\xk_{i,0}}{I_1}=\frac{\xk_{i,0}}{\textrm{tr}\left({\xk_{i,0}}\right)}=\frac{\xk_{i,0}}{\xk_{i,0,11}+\xk_{i,0,22}+\xk_{i,0,33}},
\end{eqnarray}
\noindent although any of the two other invariants would be appropriate. The mean ($\xk_e$) and variance-covariance matrix ($\xV$) of the statistical sample is defined in the usual way \cite{je1}:
\begin{eqnarray}
	\label{eq:ExpVal}\mathbf{k}_e=E(\mathbf{k}_i)=\frac{1}{N}\sum^{N}_{i=1}\mathbf{k}_i,\\
	\nonumber\label{eq:VarCov}\mathbf{V}=E((\mathbf{k}_i-\mathbf{k}_e)(\mathbf{k}_i-\mathbf{k}_e)^T)=\\
	\frac{1}{N-1}\sum^{N}_{i=1}(\mathbf{k}_i-\mathbf{k}_e)(\mathbf{k}_i-\mathbf{k}_e)^T.
\end{eqnarray}

\noindent Note, that due to normalization the number of independent quantities in $\xk_i$ equals $5$.

We assume that the elements of the mean tensor $\xk_e$ are independent random variables and that they have univariate normal distribution. Although the normality is approximate for normalized data sets, based on our experience, the error here is negligible (see the Appendix example). We investigate the closeness of the AMS and the mesotectonic stress tensors. Their nearness is not formulated as a strict equality as there are many observations that contradict such a strong relation, but it is expressed on statistical grounds. As both $\xk_e$ and $\xsigma$ are tensor valued random variables, it is argued that these two are able to mutually tighten the range of plausible principal directions. Let $E(\xA):=\{\xx\in\xR^3:\xA\xx=\lambda\xx, \left\|\xx\right\|=1\}$ denote the set of unit eigenvectors of the $3\times 3$ matrix $\xA$ with the corresponding eigenvalues $\lambda\in\xR$. The main hypothesis expresses that the eigenspaces of the two tensors are statistically indistinguishable,
\begin{eqnarray}
\label{equ:basic_assumption}
E(\xk_e)\cong E(\xsigma),
\end{eqnarray}

\noindent where sign $\cong$ denotes statistical equivalence. Our main interest is closely rotationally anisotropic data sets as close intermediate and maximal eigenvalues (i.e. nearly oblate ellipsoids, Fig. \ref{fig:00}. b) are typical for soft sediments \cite{hr}.

\section{Stochastic method for nearly isotropic tensors}
\label{sec:AMS}

The classical approach of tensor statistics assumes that the tensor is \emph{sufficiently anisotropic} \cite{he}. In this case the  confidence intervals of the eigendirections can be approximated with ellipses and can be derived analytically, so the applied \emph{linearization} leads to negligible errors. (\cite{je1} introduced this method in geosciences, since which it has been widely used.)

As both Hext and Jelinek point out, \emph{close to rotationally anisotropic tensors} cannot be evaluated by classical methods due to the non-linear dependence of the eigenvectors on the matrix elements. It is worth to mention that, even for rotationally anisotropic or isotropic tensors, three mutually orthogonal eigendirections can be computed by the widely used algorithms (let us call the later procedures \emph{direct methods}). Direct methods, in general, fail to recognize that linear combinations of the computed eigendirections might also belong to the eigenspace of the tensor. Rigorous treatment of such non-linearity has been carried out for $2\times 2$ matrices \cite{xu}. Instead of facing the even more complicated case of $3\times 3$ matrices, our method resolves the above mentioned non-linearity by performing a \emph{large number of linear investigations}. This enables simple hypothesis testing appropriate for determining eigendirections and distinguishing between eigenvalues within the data set.

\subsection{Identification of eigendirections}
By definition, the $\lambda_i$ eigenvalue and the $\xu_i$ eigenvector of the tensor $\xk_e$ fulfills
\begin{eqnarray}
	\mathbf{k}_e\mathbf{u}_i=\lambda_i\mathbf{u}_i,
	\label{eigen_def}
\end{eqnarray}

\noindent where $i\in\{1,2,3\}$, the eigenvector is normed ($\left\|\mathbf{u}_i\right\|=1$) and due to symmetry the eigenvalues $\lambda_i$ are real. Let $U$ denote a finite set of unit vectors ($\xu$) pointing to the vertices of some (more or less) regular and sufficiently fine triangulation of the unit sphere. Typically a unit vector $\xu\in U$ fails to be an eigenvector of $\xk_e$, however, based on (\ref{eigen_def}) one can define a scalar as
\begin{eqnarray}
\lambda=\mathbf{u}^T\mathbf{k}_e\mathbf{u}. 
\label{lambda_def}
\end{eqnarray}

\noindent With this in mind, the vector $\xe$ can be calculated via
\begin{eqnarray}
	\mathbf{e}=\mathbf{k}_e\mathbf{u}-\lambda\mathbf{u}=\mathbf{k}_e\mathbf{u}-(\mathbf{u}^T\mathbf{k}_e\mathbf{u})\mathbf{u}.
	\label{error_def}
\end{eqnarray}

\noindent Note, that $\left\|\mathbf{e}\right\|$ is a measure of the deviation for $\xu$ meeting eq. (\ref{eigen_def}). Our construction guarantees, that $\xe=\mathbf{0}$ iff $\xu=\xu_i$, and then $\lambda=\lambda_i$. The right-hand side of eq. (\ref{error_def}) is linear respect to the elements of the matrix $\xk_e$. We aim to decide about each elements of $U$, whether it meets to be an eigenvector of $\xk_e$. Hence, the null and alternative hypotheses of the \emph{multivariate statistical test} \cite{ti} are formulated as 
\begin{eqnarray} 
H_0: \xe=\mathbf{0},\\
\nonumber H_1: \xe\neq \mathbf{0}.
\label{T2_test_u}
\end{eqnarray}

A linear combination of normally distributed random variables is also normal, hence the elements of $\xe$ follow a normal distribution and hence we use the one-side version of \emph{Hotelling's $T^2$ } test \cite{si}. The test statistics has an $F$-distribution with parameters $p_1=2$ and $p_2=N-2$. (Detailed explanation of the method is provided in the Appendix.) All $\xu\in U$ vectors fulfilling $H_0$ are accepted as possible eigenvectors of the statistical sample, they form a subset of $U$:
\begin{eqnarray}
\tilde{U}:\{ \xu\in U\quad \left| H_0 \right. \text{ is valid }\} \subseteq U.
\end{eqnarray}

\noindent Nevertheless, acceptance criteria strongly rely on the variation of the original sample. Result of the computation can be easily visualized by marking points related to the elements of $\tilde{U}$ in a stereonet.

\subsection{Identification of eigenvalues}
As in (\ref{lambda_def}), an eigenvalue-like quantity can be computed for any unit vector. This implies, that a statistical test can be used to distinguish between significantly different eigenvalues. This test might be evaluated pairwise for all elements of $\tilde{U}$, although it seems to be more natural to compare $\lambda$ against the directly computed eigenvalues ($\lambda_i$) of $\xk_e$. It is clear from (\ref{lambda_def}) that $\lambda$ is a random variable which depends linearly on the elements of $\xk_e$, hence it follows a normal distribution. A statistical test is defined with the null and alternative hypotheses:
\begin{eqnarray} 
H_0: \lambda=\lambda_i,\\
\nonumber H_1: \lambda\neq \lambda_i,
\label{t_test}
\end{eqnarray}

\noindent where $i\in\{1,2,3\}$. As $\lambda$ is a scalar quantity, here a one-sided $t$-test is appropriate for the test statistics. If $H_0$ is valid at any value of $i$ then our data do not provide any reason to distinguish between $\lambda$ and $\lambda_i$, in other words they are \textit{indistinguishable} based on the statistical sample. The easiest way to indicate statistically different eigenvalues is a consequent coloring of the accepted eigendirections in the the above mentioned stereonet (see Figure \ref{fig:classify}).

\subsection{Statistical analysis}
\label{StatAnal}

\begin{figure}
   \includegraphics[width=1.00\textwidth]{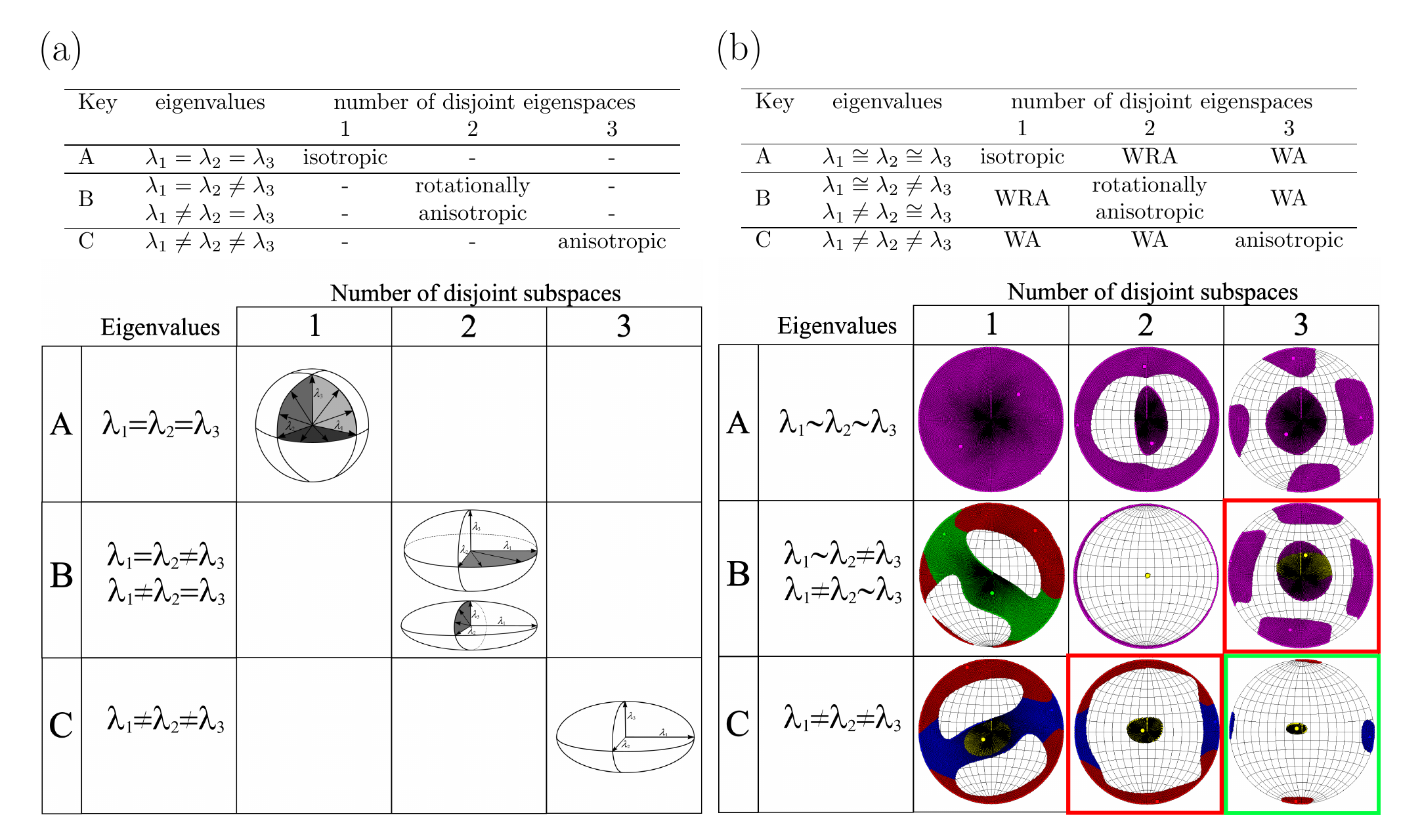}
	 \caption{Classification of $3{\times}3$ tensors with respect to the number of different eigenvalues (A-C) and disjoint eigenspaces (1-3). (a) Deterministic tensors can be associated with the ellipsoids in Fig. \ref{fig:00}. (b) In case of tensors with stochastic elements all classes are filled. The coloring of the stereograms encodes the following magnitudes of the eigenvalues: red - maximal, blue - intermediate, yellow - minimal, green: indistinguishable minimal and intermediate, purple: indistinguishable intermediate and maximal or isotropic. Abbreviations: WA: weakly anisotropic, WRA: weakly rotationally anisotropic, $\cong$ refers to the situation, when two eigenvalues are indistinguishable for a statistical test.}
		 \label{fig:classify}
\end{figure}

So far statistical tests have been introduced to identify eigendirections and identify significantly different eigenvalues from the input data. While in the case of deterministic matrices it is sufficient to investigate either the eigenvalues or the disjointness of the eigenspaces to decide about isotropy or rotational anisotropy (see Fig. \ref{fig:classify}a), stochastic tensors require both. The fact, that both the eigenvalues and the eigenspaces are needed for such an investigation, stems from the nonlinear dependence of the eigendirections and eigenvalues on the matrix elements in case of direct computation. Since the variation of the elements influence the confidence intervals of the eigendirections and eigenvalues differently, it is possible to have disjoint eigenspaces with indistinguishable eigenvalues as well as separable eigenvalues that may be accompanied by a (partially) unified eigenspace (Fig. \ref{fig:classify}b). As the simulated data sets clearly show in Figure \ref{fig:classify}b, all the possible pairing according to the number of different eigenvalues and number of disjoint eigenspaces might occur. For our later work we distinguish similar cases in the table by names: cells, which are anisotropic based on either the multiplicity of the eigenvalues or the number of disjoint eigenspaces are called \emph{weakly anisotropic} (WA) as they fail to be fully anisotropic (A3, B3, C1, and C2). Likewise, the cases which happen to fulfill exactly one of the requirements of rotational anisotropy are called \emph{weakly rotationally anisotropic}  (WRA, cases A2 and B1). The completely filled table of stochastic tensors underscores the importance of evaluation based on both the disjointness of the eigenspace and the multiplicity of the eigenvectors; the methods in the literature focusing solely on the eigenvalues are incomplete.

\section{Stress inversion in a stochastic way}
\label{sec:TECT}

Stress inversion is a synthetic term for methods used to reconstruct former stress fields by investigating observed faulting patterns of rocks. Most of the methods in the literature are based on the Wallace-Bott hypothesis \cite{bott,wall}. We are aware about the ambiguity of stress inversion methods, namely whether the stress, or the infinitesimal strain tensor is approximated by their application (e.g. comments of \cite{tw} or \cite{ga}). However, our mild constitutive assumption (see eq. \ref{equ:basic_assumption}) guarantees that the eigendirections of the infinitesimal strain and stress tensors coincide, thus this ambiguity is resolved. From the numerical point of view, each stress inversion methodology (for example \cite{ang}, \cite{ha}) sets up an optimality condition considered as the best approximation of the Wallace-Bott hypothesis for noisy input data. Even though the appropriateness of the Wallace-Bott hypothesis might be challenged on mechanical and statistical grounds \cite{li}, in this work we accept it as an adequate assumption for sediments. Instead of an arbitrary optimality condition, a stochastic approach can be argued to provide a deeper insight. It highlights fault patterns that are more probable under a given loading. By keeping a probabilistic viewpoint, a path similar to the weakly anisotropic procedure in the previous section can be followed. In other words an appropriate vector space can be sought that can be associated with the space of stress tensors pointwise (as we associated the unit-sphere with the eigendirections of $\xk_e$). 

As is well-known, the balance of angular momentum leads to the conclusion that in a fixed orthonormal basis the stress tensor $\xsigma$ can be represented by a symmetric matrix. We produce it's orthogonal diagonalization as
		\begin{eqnarray} \label{eq:orth_decomp}
		\xsigma=\xQ\xDelta\xQ^{T},
		\end{eqnarray}
where $\xQ$ is an orthogonal matrix, i.e. $\xQ^T\xQ=\xQ\xQ^T=\xI$ with $\xI$ being the identity. $\xDelta$ is a diagonal matrix with real elements (in fact, it contains the eigenvalues of $\xsigma$, also known as principal stresses). It is easy to show that for a given $\xsigma$ each $\bar{\xQ}$ with $\det(\bar{\xQ})=-1$ fulfilling eq. (\ref{eq:orth_decomp}) can be substituted with another $\xQ$, of which the determinant equals $1$ by simply multiplying one or three columns of $\bar{\xQ}$ by $-1$. As we seek eigendirections plotted on the lower hemisphere, this study is invariant under such a transformation. Hence only orthogonal matrices of real rotations are sought, i.e. we associate the space of stress tensors with the special orthogonal group $SO(3)$.

The Wallace-Bott hypothesis states that the slip direction $\xt$ (measured as a striae on the fault surface) coincides with the shear direction $\xs$ computed for $\xsigma$ at a fault plane, itself characterized by its unit normal $\xn$. It is also known \cite{ang2,si}, that the eigendirections and the shear direction are invariant under the following transformation of an arbitrary stress tensor $\xsigma_0$:
\begin{eqnarray} \label{eq:transform}
\xsigma=\alpha\xsigma_0+\beta\xI,
\end{eqnarray}
\noindent where $\alpha\in\xR\setminus\{0\}$ and $\beta\in\xR$. Since stress-inversion in its own is not sufficient to determine $\alpha$ and $\beta$ we choose the most convenient value for these parameters: for a given $\xsigma_0$ one can find a unique pair of $\alpha$ and $\beta$ such way, that the traction of $\xsigma$ coincides with the slip direction $\xt$ and consequently with the shear direction $\xs$. Whence we seek $\xsigma$ to fulfill
\begin{eqnarray} 
\label{eq:wall-bott}
\xt=\xs=\xsigma\xn.
\end{eqnarray}
Most of the other methods aim to find an optimal $\xsigma$ to explain the measured data $\xt_i$ and $\xn_i$ ($i=1...N$). Applying eq. (\ref{eq:orth_decomp}) for any fault-slip data (after multiplying by $\xQ^T$ from the left) equation (\ref{eq:wall-bott}) can be reformulated as 
\begin{eqnarray} 
\label{eq:kkk}
\xQ^T\xt_i=\xDelta_i\xQ^{T}\xn_i.
\end{eqnarray}

Observe that, for a fixed $\xQ$ and measured $\xt_i$ and $\xn_i$, the three non-zero elements in the diagonal of $\xDelta_i$ is uniquely determined. For brevity we define $\xdelta_i=\diag(\xDelta_i)$. Let us discretize $SO(3)$ with a sufficiently finite grid and associate each gridpoint with a positive integer $j\in\{1,...,M\}$. Such a discretization can be carried out by unit quaternions \cite{kui}. Our construction produces a vector of principal stresses, $\xdelta_{i,j}$ for each measurement ($i=1...N$) and each gridpoint in the discretization. Nevertheless, the principal stresses at a given gridpoint (i.e. at a fixed $\xQ$) might differ significantly as $i$ is varied. For a fixed $j=\tilde{j}$ the principal stresses can be collected for all fault-stria in a $3\times N$ matrix as
\begin{eqnarray} 
\label{eq:computed_deltas}
\xD_{\tilde{j}}=[\xdelta_{1,\tilde{j}},...,\xdelta_{N,\tilde{j}}].
\end{eqnarray}

\noindent Each row of $\xD_{\tilde{j}}$ forms a statistical sample and can be tested, that none of them has a standard deviation exceeding a given threshold $v_l$. Let us denote unit vectors in the standard basis of $\xR^3$ to $\xg_k$, where $k\in\{1,2,3\}$. Thus the test-hypothesis is formulated as:
\begin{eqnarray} 
H_0: s_N(\xg_k^T\xD_{\tilde{j}})\leq v_l \quad \textnormal{for} \quad k=1,2,3,\\
\nonumber H_1: s_N(\xg_k^T\xD_{\tilde{j}})>v_l \quad \textnormal{for any} \quad k.
\label{var_test}
\end{eqnarray}

If all three rows of $\xD_{\tilde{j}}$ exhibit an acceptably small variation (below $v_l$), then there is no reason to exclude $\xQ$ as a matrix of the eigenvectors of $\xsigma$ and $\bar{\xdelta}_{\tilde{j}}=N^{-1}\Sigma_{i=1..N}\xdelta_{i,\tilde{j}}$ as a most probable solution for its three eigenvalues. The check of the test hypothesis (which depends on the parameter $v_l$) is carried out by the properly scaled $\chi^2$ distribution (details in Appendix). As it is inherent in the method, we three mutually orthogonal directions (the columns of $\xQ$) are accepted or rejected. The three directions can then be plotted on a stereonet and colored based on the magnitudes of the elements of $\bar{\xdelta}_{\tilde{j}}$. Increasing the value of $v_l$ leads to a larger cover of the stereonet of accepted eigendirections of plausible stress tensors.

\section{Combined evaluation of AMS and mesotectonic field data}
\label{sec:JOINT}
Combined evaluation of tensor-related data sets might have different levels \cite{si}. A simple comparison of the eigendirections of the tensors can be made using standard tools of vector statistics. Such an approach has a serious shortcoming as it drops the tensor nature of the involved quantities. If the matrices representing the tensor quantities and even the covariance matrices are available, then an element-wise test for parity can be made. However such a procedure can be regarded as too strict in this case as the ellipsoids of the AMS and the stress tensor might have different eccentricities due to non-deformational reasons. In this work we intorduce a hypothesis test to confirm that the mutually orthogonal eigendirections of $\xk_e$ and $\xsigma$ are sufficiently close as it is postulated in eq. (\ref{equ:basic_assumption}). 

To reach this goal, all accepted eigendirections and eigenvalues of the mesotectonic data (encoded by the matrices $\xQ$ and  $\bar{\xdelta}_{\tilde{j}}$, respectively) are tested against the AMS data. If all the three columns of $\xQ=[\xq_1,\xq_2,\xq_3]$ can be accepted as principal directions of the magnetic susceptibility tensor and even the eigenvalues of the two tensors are plausibly close, then they can be considered to be reflecting the same deformational phase. As a hypothesis test of these requirements can be formulated as
\begin{eqnarray} 
\label{final_test}
H_0: \xk_e\xq_k-(\xq_k\xk_e\xq_k)\xq_k=\mathbf{0},\\
\nonumber H_1: \xk_e\xq_i-(\xq_i\xk_e\xq_i)\xq_i\neq\mathbf{0}.
\end{eqnarray}

Observe the linearity of these expressions: as $\xQ$ is fixed by the discretization of $SO(3)$, the random variables of the above test are $\xk_e$ and $\bar{\xdelta}_{\tilde{j}}$. As shown in the Appendix, Hotelling's $T^2$ squared test is used for (\ref{final_test}). As a byproduct, $T_0$ and the maximal value of $T_2$ (as $\xQ$ is varied) can be used for characterize the closeness of the principal directions via
\begin{eqnarray} 
C=\frac{T_0-T_2}{T_0}.
\label{eq:c_value}
\end{eqnarray}

The value of $C$, by definition, is smaller than one. If it was negative, then the two data sets express different tensors thus there is no reason to assume a common origin. For positive $C$ a common (deformational) origin of the AMS and the fault-slip data is probable, higher values hint at even better agreement between the two data sets. The applicability of the new method is illustrated by seven field examples in the next Section. A custom-made algorithm in MATLAB was implemented for the computations. It uses several subroutines for the visualization of stereonets from \cite{alm}.

\section{Field examples}
\label{sec:EXAMP}

Although the data presented below are of extensional or strike-slip types (as sediments of the Pannonian Basin were dominantly deformed by extension or transtension), the method can be predicted to be readily applicable to compressional stress fields situations. After having applied a tilt-test, in all cases examined here, the deformation registered by the magnetic fabric occurred early in the deformation history, i.e., while in a sub-horizontal bed position. Therefore they can be compared to early faults and related stress axes that also formed before tilting. To provide a detailed view, the entire fault-slip data set for two of our examples in the Appendix are presented. This shows the deformation history involving tilting of the sedimentary beds and it also demonstrates that we are only dealing with the earliest brittle deformation event which affected the studied outcrops. 

\begin{table}[!h]
\caption{Geographical data for the analyzed field examples. The source of the first published AMS and mesotectonic evaluation are also listed for each site.}
\label{tab:results}
\begin{adjustbox}{max width=\textwidth}
\begin{tabular}{@{}lllllllll}
Site      & Country & X                   &Y	                  & Age		    & Rock &AMS-source &Evaluation source & Field data \\\hline  
Cezlak       & SLO  &$46^\circ 25'13.07''$&$15^\circ 26'18.70''$& E.Miocene & granodiorite &this work & \cite{fo0} &Vrabec\\
Ro\v{s}poh      & SLO  &$46^\circ 36'28.15''$&$15^\circ 36'10.41''$& E.Miocene & siltstone&this work & this work &Fodor, Vrabec, Jelen\\
Lovrenc      & SLO  &$46^\circ 33'7.42''$ &$15^\circ 24'45.1''$ & E.Miocene & siltstone   &this work & \cite{fo} &Fodor, Jelen, Trajanova\\
Feny\H of\H o& HU   &$47^\circ 22'14.67''$&$17^\circ 47'58.40''$& Eocene    & clay &this work & \cite{ma3} &Fodor\\
\'Obarok     & HU	  &$47^\circ 30'34.55''$&$18^\circ 34'3.43''$ & Oligocene & clay & \multicolumn{2}{l}{\cite{sib}} &Fodor\\
S\'aris\'ap  & HU	  &$47^\circ 40'0.63''$ &$18^\circ 40'16.12''$& Oligocene & siltstone &this work & \cite{sib} &Bada, Fodor, Maros\\
Pesnica      & SLO  &$46^\circ 36'33.75''$&$15^\circ 39'51.11''$& E.Miocene & marl &this work & \cite{ma4} &Fodor, Vrabec, Jelen\\
\end{tabular}
\end{adjustbox}
\end{table}

All of our calculations were carried out at the usual $\alpha=0.05$ significance level, the geographical data are shown in Table \ref{tab:results}. The classical AMS plots were obtained by Anisoft 4.2. \cite{ch}; and the method of Angelier \cite{ang,ang2} was used for stress tensor calculations. For the $v_l$ parameter of the stress-inversion method $v_l=1.5$ was taken, in the case of an extensional field, and $v_l=2.0$ for strike-slip fields. (An accepted result of any stress-inversion method can be used to determine a plausible value for $v_l$, c.f. Appendix.) Beyond the stereonets, the value of $C$ (defined in the previous Section) was calculated for all examples. 
Furthermore, the maximal extension of the confidence intervals of the $K1$ direction were determined using both the classical AMS and the new combined evaluation methods. In detail, the double of the semi-major axis $E_{12}$ was computed using the classical method and then compared to the furthest angular distance between accepted eigendirections belonging to the maximal eigenvalue in the combined evaluation. This latter angle is denoted by $\psi$. These results are in Table \ref{tab:results2}., a step by step presentation of the method is given in the Appendix for one of our examples (Feny\H{o}f\H{o}).

\begin{figure}
		\includegraphics[width=0.95\textwidth]{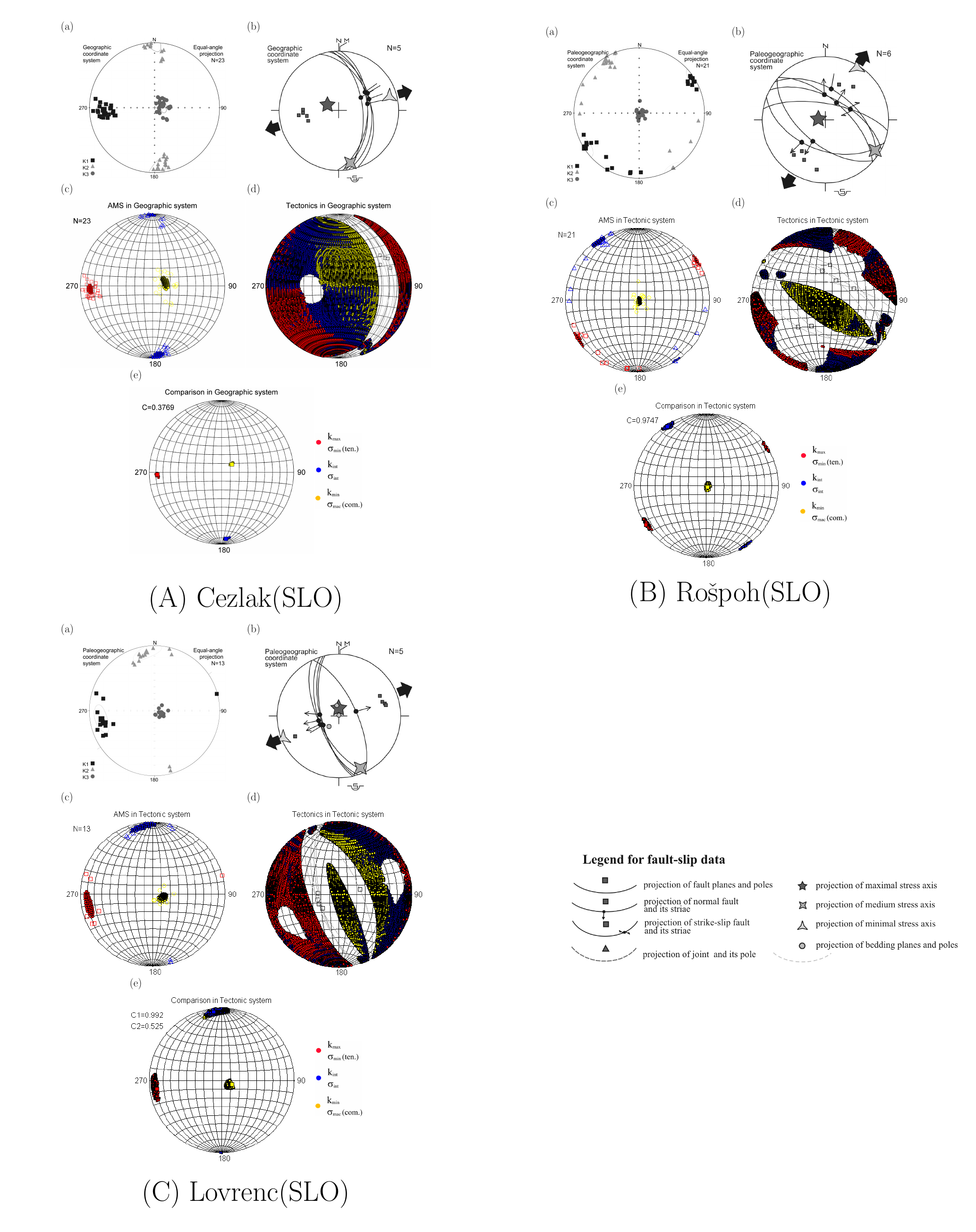}
		\caption{AMS data, main axes, fault-slip data and calculated stress axes for the sites Cezlak (A), Ro\v{s}poh (B) and Lovrenc (C), Slovenia. The panels for each site are AMS with classical approach (a) and with the novel method (c); stress axes calculated by the method of \cite{ang} (b) and the novel method (d), respectively. Comparison of commonly obtained AMS and stress axes (e). For coordinates, data source see Table \ref{tab:results}. Goodness of the fit for the three sites: $C_{\text{Cezlak}}=0.3769$, $C_{\text{Ro\v{s}poh}}=0.9747$, $C_{\text{Lovrenc}}=0.9920$.}
		\label{fig:03}
\end{figure}

A benchmark-like test is given based on Cezlak, Slovenia (Figure \ref{fig:03}A). Even though this is a magmatic rock, the additional information about strain makes it a perfect example to introduce the new procedure. Here the K1 direction of AMS is parallel to the strain markers observable in the field and under the microscope, while the markers of brittle deformation are weak \cite{ma5}. The formation is made of granodiorite, which suffered ductile deformation at an estimated temperature of $400-450 ^\circ$C followed by brittle deformation after cooling \cite{fo0}. This rock has a high susceptibility ($\approx 10^{-2}$ SI), extremely high degree of AMS ($\approx 35\%$ in average) and lineation ($\approx 20\%$ in average) \cite{ma}. In this case the classical method by V. Jelinek is a perfect procedure to determine the orientation of the AMS ellipsoid. Observe, that the confidence ellipses of the classical and the new solutions overlap precisely. Although the small number of fault-striae make the stress-inversion uncertain, it nonetheless reflects an extensional stress field. The possible principal directions calculated from the markers of brittle deformation cover almost the entire stereonet underscoring the insufficient number of measurements. Finally, the simultaneous evaluation not only selects a few solutions from the vast orthonormal bases in the mesotectonic side, but it also tightens the region of acceptance for the AMS measurements.

As it was mentioned earlier, the real targets of the proposed method are sediments with low degree of magnetic susceptibility and even lower degree of lineation such as the data set from Ro\v{s}poh, Slovenia (Figure \ref{fig:03}B). For this first example locality the susceptibility is weak ($\approx 10^{-4}$ SI) and accompanied by moderate anisotropy ($\approx 7.6\%$) and weak lineation ($\approx 1\%$). In terms of AMS the new method yields an identical solution with the classical method. Brittle markers on  conjugated faults reflect an extensional stress field. The loose definition of the extensional direction in the mesotectonic data is also reflected well in the evaluation of the new method, however the combined evaluation with the AMS narrows down the direction of the extension.

In the case of Lovrenc, Slovenia (Figure \ref{fig:03}C, susceptibility $3.5\cdot 10^{-4}$ SI, anisotropy $9.6\%$, lineation $0.7\%$) the number of AMS data is lower. A small number of faults represent the first deformation event that occurred in horizontal bed position. Comparison against the mesotectonic data underscores the directions suggested by the AMS stereonet. 

\begin{figure}
		\includegraphics[width=0.95\textwidth]{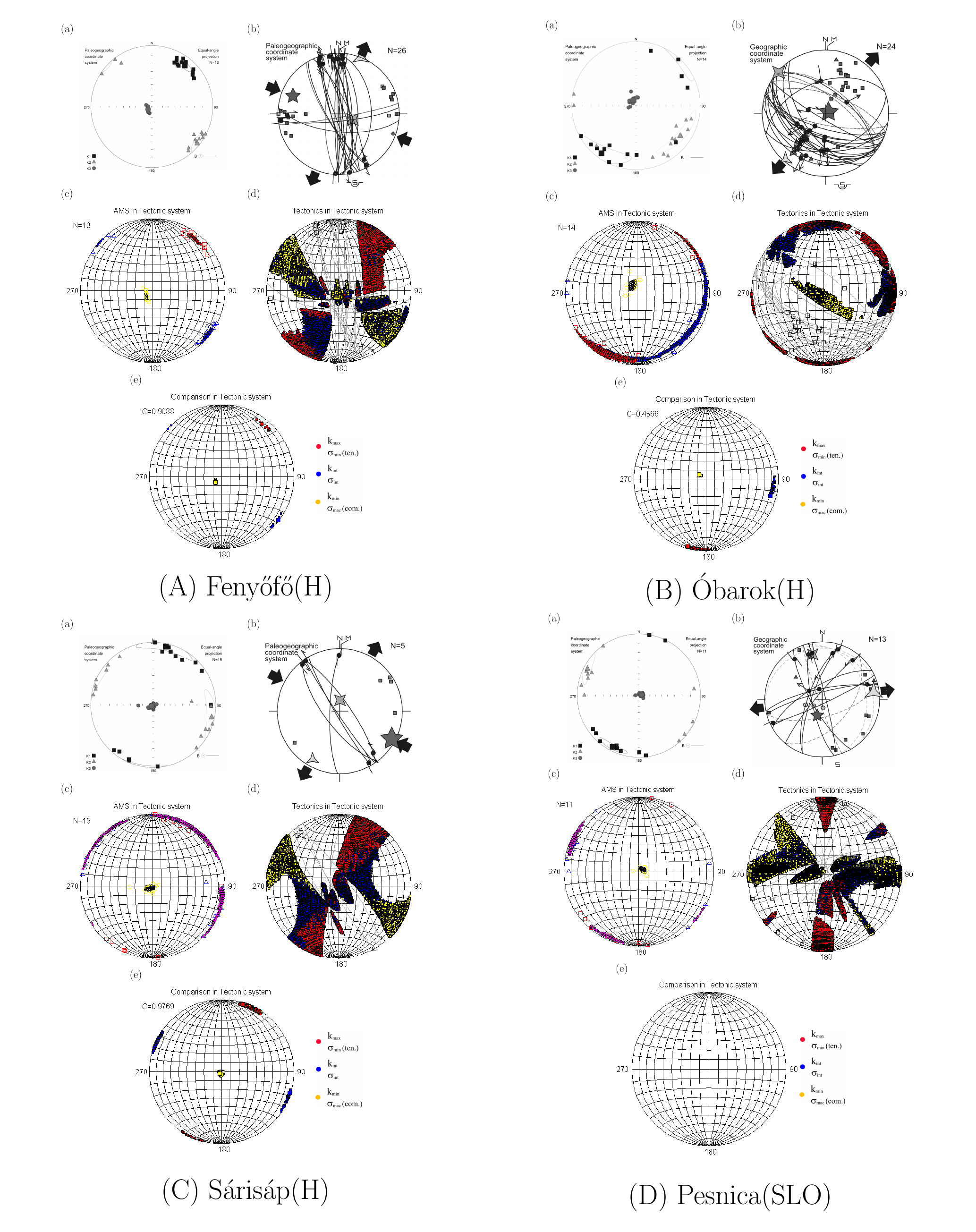}
		\caption{AMS data, main axes, fault-slip data and calculated stress axes for the sites Feny\H of\H o (A), \'Obarok (B) S\'aris\'ap (C), Hungary and Pesnica (D), Slovenia. The panels for each site are AMS with classical approach (a) and with the novel method (c); stress axes calculated by the method of \cite{ang} (b) and the novel method (d), respectively. Comparison of commonly obtained AMS and stress axes (e). For coordinates, data source see Table \ref{tab:results}. Goodness of the fit for the four sites: ${C_{\text{Feny\H of\H o}}=0.9088}$, ${C_{\text{\'Obarok}}=0.4366}$, ${C_{\text{S\'aris\'ap }}=0.9769}$, ${C_{\text{Pesnica}}=0.0000}$.}
		\label{fig:04}
\end{figure}

The next example is from Feny\H of\H o, Hungary (Figure \ref{fig:04}A, susceptibility $3.5\cdot 10^{-4}$ SI, anisotropy $9.6\%$, lineation $0.7\%$). At this locality measurements tightly constrain the direction of the AMS. The new method leads to a similar outcome to the classical method. The mesotectonic data reflect a well-defined pattern of a strike-slip type deformation. Joint evaluation enhances the precision of the extensional direction. Note also that the two, well-defined data sets indeed reflect the same stress field.

The AMS measurements for \'Obarok, Hungary (Figure \ref{fig:04}B, susceptibility $1.6\cdot 10^{-4}$ SI, anisotropy $7.9\%$, lineation $0.8\%$) represent a case where the maximum and intermediate directions exhibit a rather large scatter. Although there is no overlap between the two populations, such an extended confidence interval ($~40^\circ$) is not appropriate for the classical method. Evaluating with the new method produces an overlapping set which demonstrates that the classical method significantly underestimate the confidence regions in this case. Weakly anisotropic (C2 type) AMS data hint at no extensional direction. Despite the large number of mesotectonic markers, tensional direction determined from fault-slip data also have considerable uncertainty. The combined evaluation reveals a clear extensional direction in NNE-SSW. 

The other weakly anisotropic (3B type) type occurs in case of S\'aris\'ap, Hungary (Figure \ref{fig:04}C, susceptibility $4.7\cdot 10^{-4}$ SI, anisotropy $10.2\%$, lineation $0.8\%$) in the horizontal plane of the classical AMS diagram two clusters are obvious. However, there is an indication of uncertainty for the character of the axes of the ellipsoid: one maximum falls in the dominant intermediate directions, consequently one intermediate direction is associated with the other maxima. The new method reveals that this uncertainty is indeed significant: statistically there is no hint of which cluster represents the maximal or the intermediate direction. This example also demonstrates that any acceptance of data sets solely based on their confidence intervals (which can be even tighter, than in present case) is not reliable: it is advisable to check the clusters on the stereograms. Lukily, a few mesotectonic markers constrain a strike-slip stress field, which is reflected as a narrow ranged extensional direction computed by the new method. The combined evaluation shows that the stress field has a definite extensional direction that is close to the classical mesotectonic evaluation.

The final example is from Pesnica, Slovenia (Figure \ref{fig:04}D, susceptibility $1.7\cdot 10^{-4}$ SI, anisotropy $6.2\%$, lineation $0.3\%$). Here the AMS ellipsoid is closer to a rotationally anisotropic type than in the previous examples. At first sight it is similar to S\'aris\'ap (Figure \ref{fig:04}) as the resultant susceptibility is of the 3B type: in the magnetic foliation plane two populations are clearly distinguished, but it is impossible to say, which is the population of maxima and that of the intermediate directions. The orientation of the stress field is well constrained by the mesotectonic markers, as is confirmed by the new method. The combined evaluation provides an empty stereonet which means, that either the AMS is not of deformational origin or at least it only reflects, very weakly, an earlier deformational phase than those suggested by the fault-slip data. 

\begin{table}
\caption{Comparison of the maximal confidence interval of the extensional direction by the classical and the combined methods. $v_l$ is the threshold parameter of acceptance in the stochastic stress inversion, $2E_{12}$ is the semi-major diameter of the confidence-ellipse in the classical AMS procedure, $\psi$ is the maximal angular distance between two accepted eigenvectors which both belong to the maximal eigenvalue in the combined evaluation. $2E_{12}$ and $\psi$ both measure the extent of the confidence region. Note that the combined evaluation results in tighter confidence intervals in all cases (except Pesnica where it is meaningless). It is especially powerful in case of nearly rotationally AMS (\'Obarok and S\'aris\'ap).}
\label{tab:results2}
\begin{tabular}{l|c|c|c|c}
Locality           & $v_l$ & Trend of extension &$2E_{12}$      & $\psi$ \\\hline          
Cezlak, SLO        & 1.5   &$86-266^\circ$   &$14.8^\circ$			& $4.9^\circ$ \\
Ro\v{s}poh, SLO       & 1.5   &$58-238^\circ$   &$19.4^\circ$	    & $11.7^\circ$ \\
Lovrenc, SLO       & 1.5   &$82-262^\circ$   &$30.4^\circ$			& $21.5^\circ$ \\
Feny\H of\H o, H   & 2.0   &$42-222^\circ$   &$18.2^\circ$			& $10.6^\circ$ \\
\'Obarok, H 		 	 & 1.5   &$15-195^\circ$ 	 &$62.4^\circ$			& $15.2^\circ$ \\
S\'aris\'ap, H		 & 2.0   &$13-193^\circ$   &$57.4^\circ$			& $14.8^\circ$ \\
Pesnica, SLO		   & 2.0   &-   	  &-				   				& -
\end{tabular}
\end{table}

\section{Discussion}
The field examples illustrate the power of the new method using a statistical approach. On one hand, nearly rotationally anisotropic AMS data sets with high confidence angles can be evaluated reliably, as the method is based on a new, linearization-free technique. It extends the classical method into this regime. On the other hand, stress-inversion is also carried out statistically, enabling a hypothesis test for the degree of coincidence of the AMS and stress tensors. 

The idea of combination of AMS with mesotectonic data for sediments is not new -- generally axes K1 and S3 tend to have similar orientations. This suggests that the two techniques depict the same deformation, as pointed out in several examples and in the literature \cite{ci}. A slight temporal difference might have existed between grain-scale and mesoscale deformation, because the AMS pattern could be imprinted in relatively soft status of sediments, prior to the progressive lithification events that are a pre-requisite to brittle faulting (without lithification, most of the studied rocks would show deformation bands, not faults and joints). However, tilt test of fractures clearly indicate that the extensional direction (S3 axis) was deduced from the earliest mesoscale deformation events. It is clear that the new method may not be sensitive enough to demonstrate differences between the strain and AMS axes, as indicated by theoretical approaches \cite{hae,jez}. However, considering the small amount of deformation, and the lack of pronounced shear zones, the coaxial nature of deformation seems highly probable. 

Nevertheless, this uncertainty might have introduced errors into the analysis. The comparison of the two data sets (AMS and fractures) seems to suggest that - on a statistical grounds - the obtained extensional axes cannot be separated. This similarity may give grounds for thinking that such comparison might have value and could be used for refined analysis of deformation in weakly deformed sediments. In addition, the common treatment of AMS and fault-slip data by the new method facilitates a tighter range for the extensional direction (example of Feny\H{o}f\H{o}) than that calculated by the traditional, separated evaluation of AMS and stress, respectively. When the AMS lineations are well developed but the mesotectonic markers do not constrain the extensional direction precisely (examples Ro\v{s}poh and Lovrenc), the combined data set may help to better constrain the latter - if the previous conclusion about similarity is taking into account. Moreover, a more precise extensional direction can be defined when both the AMS and fault-slip data issued ill-defined axes (examples \'Obarok, S\'aris\'ap). 

Finally, as the example from Pesnica shows, it is possible to exclude a common origin for the AMS and mesotectonic markers. While in the previous examples it is highly probable that AMS and the mesotectonic markers originated from the same stress-field, then in the case of Pesnica such a possibility can be excluded. There are two options: either the AMS is not of deformational origin or the stress fields imprinting the magnetic fabric and causing the brittle deformation are not coeval.

\section{Conclusions}

In this paper a novel stochastic procedure for combined evaluation of AMS and mesotectonic data is presented. This method has a general application for the study of the ductile-brittle transition of rocks; it is particularly useful, when the AMS and mesotectonic observations come from weakly deformed soft sediments. The reason is that the AMS fabric of poorly cemented sediments tend to be nearly rotationally anisotropic and the mesotectonic markers are limited in number and quality. The new method in AMS evaluation is a perfect extension of the classical methods. Stress inversion methods in the literature for evaluating mesotectonic data operate on arbitrary optimality conditions. In the work presented here the standard methods are substituted by a stochastic approach which provides not only the principal directions, but also the statistical information needed for the combined evaluation. The hypotheses tests based on the two methods are recommended because they enhance the precision of the determination of the extensional direction of the stress field and in the same time able to recognize cases, where the AMS may not be of deformational origin or the AMS lineation and the extension direction derived from the mesotectonic data do not belong to the same tectonic regime.

\section*{Acknowledgment}
We are indebted to  M. Mattei, J. Je\v{z}ek and for their comments and suggestions, concerning the earlier version of the paper and the referees, C. Talbot and an anonymous reviewer, for their comments which significantly improved the manuscript. We thank D.H. Tarling for greatly improving the English of the paper. K. Sipos-Benk\H{o}, M. Vrabec, B. Jelen, H. Rifelj, M. Trajanova contributed to field measurements and fault-slip data evaluation.The research was supported by the Hungarian Scientific Research Fund Grant K105245 and by the J\'anos Bolyai Research Scholarship of the Hungarian Academy of Sciences [SA]

\appendix
\section{Detailed derivation of the statistical tests}
\subsection{Hypothesis test for the AMS data}
	
The linear equation (\ref{error_def}) might be written as

\begin{eqnarray}
	\mathbf{e}=\mathbf{A}+\mathbf{B}\mathbf{\hat{k}}_e, 
	\label{error_def2}
\end{eqnarray}

\noindent where $\mathbf{\hat{k}}_e$ is a vector containing the independent elements of the susceptibility tensor $\mathbf{k}_e$. (In particular, without normalization $\mathbf{\hat{k}}_e=[k_{e,11}, k_{e,22}, k_{e,33}, k_{e,12}, k_{e,23}, k_{e,13}]^T$ and with normalization we have $\mathbf{\hat{k}}_e=[k_{e,22}, k_{e,33}, k_{e,12}, k_{e,23}, k_{e,13}]^T$.) $\xA$ and $\xB$ in the above equation are determined by the components of $\xu=[u_1,u_2,u_3]^T$ (note that $\left\|\xu\right\|=1$). In case of data sets without normalization $\xA=[0,0,0]^T$ and the $3\times 6$ matrix is 

\begin{eqnarray}
	\label{eq:B_nn}
	\mathbf{B}=
	\begin{bmatrix}
			 u_1-u_1^3 &     -u_1^2u_2 & -u_1^2u_3 \\
			 -u_1u_2^2 &     u_2-u_2^3 & -u_2^2u_3 \\
			 -u_1u_3^2 &     -u_2u_3^2 & u_3-u_3^3 \\
	-2u_1^2u_2+u_2 & u_1-2u_1u_2^2 & -2u_1u_2u_3 \\
	 -2u_1u_2u_3   & u_3-2u_2^2u_3 & u_2-2u_2u_3^2 \\ 
		u_3-2u_1^2u_3  &   -2u_1u_2u_3 & u_1-2u_1u_3^2
	\end{bmatrix}^T.
\end{eqnarray}

For normed data sets they are

\begin{eqnarray}
	\label{eq:A_norm}
	\xA=
	\begin{bmatrix}
	u_1-u_1^3,-u_1^2u_2,-u_1^2u_3
	\end{bmatrix}^T,\\
	\label{eq:B_norm}
	\mathbf{B}=
	\begin{bmatrix}
	 -u_1u_2^2-u_1+u_1^3 & u_2-u_2^3+u_1^2u_2  & -u_2^2u_3+u_1^2u_3 \\
	-u_1u_3^2-u_1+u_1^3 & -u_2u_3^2+u_1^2u_2  &  u_3-u_3^3+u_1^2u_3\\
	-2u_1^2u_2+u_2      & u_1-2u_1u_2^2       & -2u_1u_2u_3 \\
	 -2u_1u_2u_3        & u_3-2u_2^2u_3       & u_2-2u_2u_3^2 \\
	 u_3-2u_1^2u_3      & -2u_1u_2u_3         & u_1-2u_1u_3^2
	\end{bmatrix}^T.
\end{eqnarray}

We remark, that the rank of $\xB$ equals 2, in other words, one of the three elements of $\xe$ is linearly dependent on one of the other two elements. In the statistical test that element (and the corresponding rows in $\xA$ and $\xB$, respectively) should be deleted. The adjusted objects are denoted to $\hat{\mathbf{e}}$, $\hat{\mathbf{A}}$ and $\hat{\mathbf{B}}$, respectively. It means, the $\xW$ variance-covariance matrix used for the test statistics is $2\times 2$, formally it is obtained via $\xW=\hat{\xB}\xV\hat{\xB}$. The test statistics for Hotelling's $T^2$ is obtained as 
\begin{eqnarray}
\label{eq:T2}
T^2=N\hat{\mathbf{e}}^T \mathbf{W}^{-1} \hat{\mathbf{e}},   
\end{eqnarray}

\noindent which (based on the above explanation of rank-deficiency) follows the $F$-distribution with parameters $p_1=2$ and $p_2=N-2$ at the $\alpha$ significance level. For the test hypothesis in eq. (\ref{T2_test_u}) we need to rescale the inverse of the $F$-distribution as 
\begin{eqnarray}
\label{eq:T0}
T_0=2(N-1)/(N-2)F_{1-\alpha,2,N-2}.
\end{eqnarray}                           

For $T^2<=T_0$ there is no reason to reject $H_0$ in \ref{t_test}, otherwise $H_1$ is accepted.

\subsection{Hypothesis test for the mesotectonic data}
In equation (\ref{eq:kkk}) $\xQ$ is a fixed orthogonal matrix (an element from the discretization of $SO(3)$), $\xn_i$ and $\xt_i$ are a measured fault and stria pair ($i=1..N$). For each measured fault-stria pair the elements of the vector $\xdelta_{i,j}=\diag(\xDelta)=[\delta_{i,j,1},\delta_{i,j,2},\delta_{i,j,3}]^T$ are computed. For fixed $j=\tilde{j}$ and $k=\tilde{k}$ $\delta_{i,\tilde{j},\tilde{k}}$ is assumed to follow normal distribution. We fix a parameter $v_l$ as a threshold of accepted variance. A known stress-inversion solution can be used to fix $v_l$, see the example below. The test statistics is computed as
\begin{eqnarray}
s_N(\delta_{i,\tilde{j},\tilde{k}})=(N-1)\left(\sigma(\delta_{i,\tilde{j},\tilde{k}})/v_l\right)^{1/2},
\end{eqnarray}

\noindent where $\sigma$ is the corrected sample standard deviation. As we carry out an upper one-tailed test, it follows a $\chi^2$ distribution with $(N-1)$ degrees of freedom at the $\alpha$ significance level. If 
\begin{eqnarray}
s_N(\delta_{i,\tilde{j},\tilde{k}})>\chi^2_{1-\alpha,N-1} 
\end{eqnarray}

\noindent holds for any $k$, then the $H_0$ hypothesis in (\ref{var_test}) is rejected.

\subsection{Hypothesis test for the combined evaluation}
Let $\xq_1$, $\xq_2$ and $\xq_3$ denote the directions of the maximal, intermediate and minimal tensile stresses, respectively. Nevertheless, each of these vectors is one the columns for $\xQ$. These vectors are orthogonal unit vectors of the $S^2$ sphere, thus they are not independent. We define $\xq=[\xq_1^T \xq_2^T \xq_3^T]^T$. Following eq. (\ref{error_def2}) a vector $\hat{\xe}$ can be defined to express the deviation from $\xq$ being an orthonormal basis of eigenvectors. Similarly to definitions (\ref{eq:A_norm}) and (\ref{eq:B_norm}) a system matrix $\hat{\xB}$ and a vector $\hat{\xA}$ can be derived by the elements of $\xq$ to fulfill $\hat{\xe}=\hat{\xA}+\hat{\xB}\xk_e$. Neglecting the linearly dependent rows of $\hat{\xB}$ (and consequently $\hat{\xA}$) one arrives to a Hotelling's $T^2$ test with parameters $p_1=3$ and $p_2=N-3$. Formally the test is given by equations (\ref{eq:T2}) and (\ref{eq:T0}).

\section{A complete fault-slip analysis of two sites}
 \begin{figure}
		\includegraphics[width=1.00\textwidth]{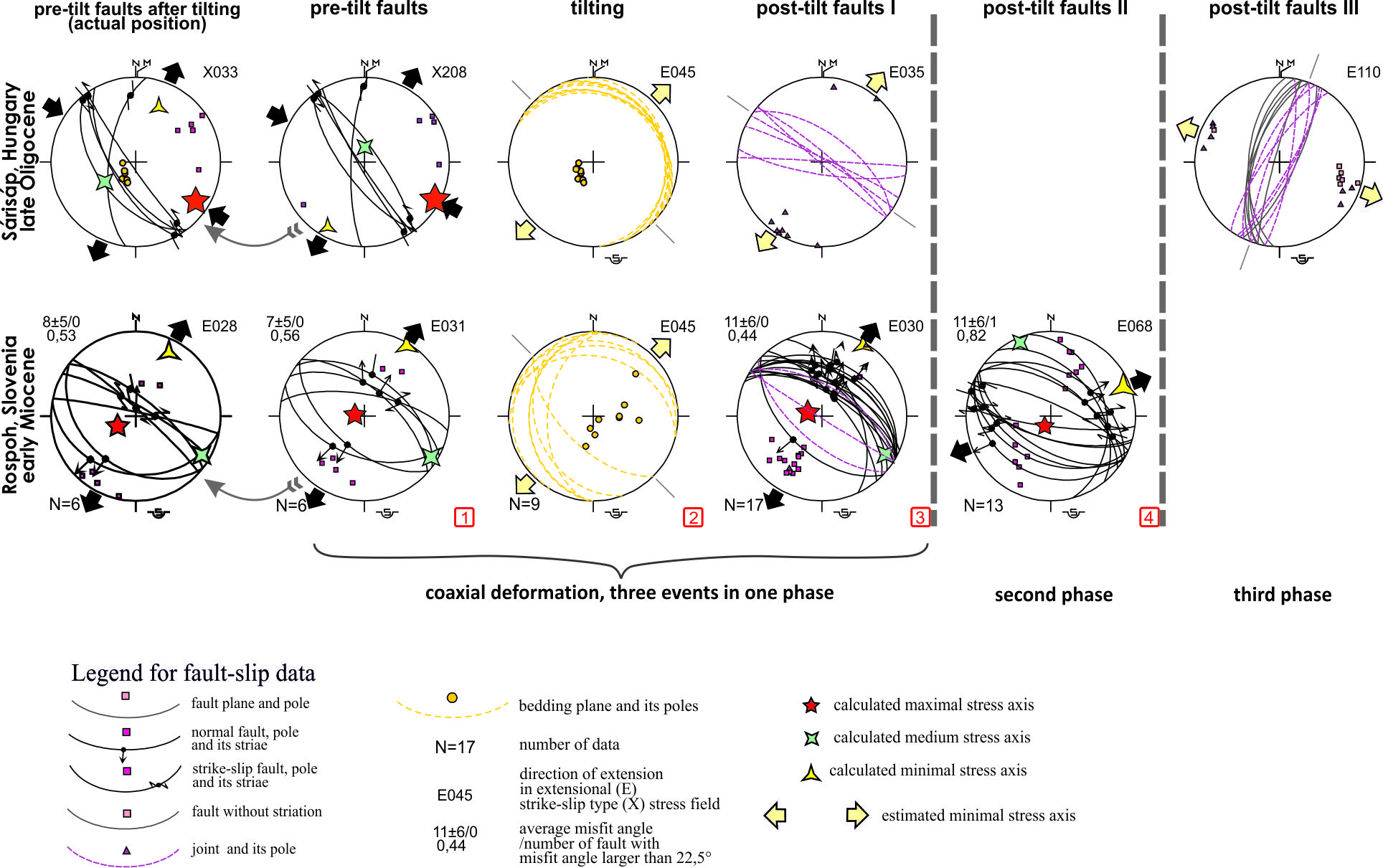}\\
		\caption{Stress field evolution in Ro\v{s}poh and S\'aris\'ap. Note the tilt test for faults (first two columns on the left side), coaxial deformation events (within one phase) and successive faulting phases with clockwise rotating stress axes. Numbers at right bottom corners indicate relative chronology of events.}
		\label{fig:complete}
\end{figure}

Although this paper does not aim to analyze the deformation history of the studied sites, we briefly present two localities with a complete fault-slip data set (Figure \ref{fig:complete}). In both sites the tilt of layers were preceded by brittle faulting, because the tilt test (left side of Figure \ref{fig:complete}) shows that the fault set is more symmetrical to the sub-vertical plane at sub-horizontal bed position than today (after tilting). This first episode of deformation was followed by the tilt itself. In Ro\v{s}poh, the variable dip direction is due to drag folding near the measured normal faults. The stress field, responsible for the tilting event can only be estimated and not properly calculated. After the tilt, normal faults (Ro\v{s}poh) and joints (S\'aris\'ap) could be formed in the same extensional stress field than the pre-tilt faults. The three events can be considered to belong to one tectonic phase. Regional analysis \cite{fo00} shows that this was the main rifting phase of the Pannonian basin. This phase was followed by a slightly different extension in Ro\v{s}poh, while it was not observed in S\'aris\'ap. On the other hand, in this latter site, a markedly different, ESE-WNW extension induced the formation of joints and small faults. This third phase can be considered as the post-rift phase of the Pannonian basin \cite{fo00,sib}. This evolution occurred during the Miocene, during the progressive burial of the studied sediments. During our analysis, only the first increment of deformation, the pre-tilt faulting was compared to AMS data.

\section{Detailed example of application}
In this appendix we provide the detailed computational results for one of our examples: Feny\H{o}f\H{o} (Figure \ref{fig:04}A). The AMS measurements consist of 13 samples, their data are given in table \ref{tab:ex_01}.

The $\xk_i$ ($i=1..13$) susceptibility tensors are computed based on the 15 directions and calibration coefficients of the KappaBridge tool. For each measurement normalization is carried out (eq. \ref{eq:normalization}) Applying equations (\ref{eq:ExpVal}) and (\ref{eq:VarCov}) we get the following mean and variance-covariance matrices:

\begin{equation}
\xk_e=
\begin{bmatrix}
 0.3434  &  0.0009 &   0.0029\\
 0.0009  &  0.3427 &   0.0037\\
 0.0029  &  0.0037 &   0.3139\\
\end{bmatrix}
\end{equation}

\begin{equation}
\xV=10^{-5}\cdot
\begin{bmatrix}
    0.1251 & -0.1507 &  -0.0178 &   0.0029 &   0.0833 \\
   -0.1507 &  0.2345 &  -0.0048 &  -0.0543 &   0.0222 \\
   -0.0178 & -0.0048 &   0.0278 &   0.0287 &  -0.0967 \\
    0.0029 & -0.0543 &   0.0287 &   0.0628 &  -0.1460 \\
    0.0833 &  0.0222 &  -0.0967 &  -0.1460 &   0.6928 \\
\end{bmatrix}
\end{equation}

Observe that the standard deviation of the elements along the main diagonal is approximately $\sqrt{1.5\cdot10^{-4}}\approx 0.00125$, which compared to the mean values around $0.333$ can be regarded as small. 

These matrices are used in the hypothesis test formulated in eq. (\ref{T2_test_u}), which produces the c) subfigure in Fig. \ref{fig:04}A. The test in eq. (\ref{t_test}) is used to color the figure. The discretization of the lower hemisphere is obtained as the intersections of equally spaced $N_1=50$ latitude lines and $N_2=200$ longitudinal lines.

The measured fault-stria data (altogether 19 measurements) is collected in Table \ref{tab:ex_02}. For the computation of the plausible stress tensor one has to define the variation limit $vl$ to apply the test in eq. (\ref{var_test}). One way of choosing this parameter is taking a result (i.e $\xQ$) of a traditional stress-inversion method and determine standard deviations for each row for eq. (\ref{eq:computed_deltas}). Either the maximum or the average of the variations are good candidates for $v_l$. In our case the method of Angelier (depicted on part b) of Figure \ref{fig:04}A) determined $291^\circ/20^\circ$ for the maximal, $112^\circ/70^\circ$ for the intermediate and $21^\circ/0^\circ$ for the minimal stress after tilting. With these directions in hand the computed eigenvalues in eq. (\ref{eq:computed_deltas}) have a standard deviation as $v_l=3.87$ in average. To keep consistence with the other sites presented in the paper a more strict, $v_l=2.00$ threshold is applied in the stochastic stress inversion. For the discretization of SO(3) altogether $260 000$ points (i.e. different $\xQ$ rotation matrices) are investigated, and at $v_l=2.0$ about $20 000$ are accepted as plausible explanation of the measured data, these are plotted in the d) part of Figure \ref{fig:04}A.

Finally, the accepted directions are tested against the AMS data as it is given in eq. (\ref{final_test}). Accepted directions are plotted in the e) part of Figure \ref{fig:04}A and finally the $C$-value of fit is computed (eq. \ref{eq:c_value}).

\begin{table}
\caption{measured AMS data of Feny\H{o}f\H{o}. Notations: s: sign of the sample o: orientation in degrees d: dip in degrees h: magnetization of the sample holder r: range of the measurement f: data in the 15 directions}
\tiny
\label{tab:ex_01}
\begin{tabular}{ r r r r r r r r r r r r r r}
\hline
	s &7913n1&7914n1&7915an1&7918an1&7919an1&7915bf1.350&7916af2.350&7916bn1&7917f1.150&7918bf1.150&7919bf1.150&7920n1&7921n1\\ \hline
	o & 70 & 78 & 70 & 71 & 83 & 70 & 78 & 78 & 79 & 71 & 83 & 82 & 82 \\ 
	   &74 & 70 & 70 & 78 & 83 & 70 & 71 & 71 & 85 & 78 & 83 & 82 & 76  \\ \hline
	d & 360 & 360 & 360 & 360 & 360 & 360 & 360 & 360 & 360 & 360 & 360 & 360 & 360 \\ 
		& 0 & 0 & 0 & 0 & 0 & 0 & 0 & 0 & 0 & 0 & 0 & 0 & 0 \\ \hline
	h & -111 & -111 & -111 & -111 & -110 & -112 & -113 & -112 & -112 & -112 & -112 & -112 & -110 \\ \hline
	r & 4 & 4 & 4 & 4 & 4 & 3 & 3 & 3 & 3 & 3 & 3 & 3 & 4 \\ \hline
	f & 760 & 709 & 711 & 698 & 747 & 1613 & 1562 & 1525 & 1497 & 1750 & 1816 & 1599 & 736 \\ 
	& 765 & 704 & 712 & 681 & 740 & 1635 & 1505 & 1496 & 1471 & 1702 & 1794 & 1604 & 716 \\ 
	& 732 & 680 & 682 & 660 & 713 & 1558 & 1474 & 1460 & 1423 & 1666 & 1731 & 1539 & 698 \\ 
	& 759 & 710 & 711 & 696 & 745 & 1614 & 1563 & 1526 & 1497 & 1750 & 1815 & 1598 & 738 \\ 
	& 762 & 702 & 711 & 677 & 738 & 1636 & 1501 & 1495 & 1472 & 1701 & 1793 & 1602 & 717 \\ 
	& 780 & 722 & 729 & 710 & 767 & 1649 & 1596 & 1563 & 1540 & 1784 & 1869 & 1659 & 747 \\ 
	& 787 & 728 & 732 & 709 & 771 & 1683 & 1572 & 1558 & 1546 & 1781 & 1876 & 1670 & 746 \\ 
	& 788 & 731 & 739 & 712 & 771 & 1677 & 1593 & 1574 & 1545 & 1789 & 1876 & 1669 & 750 \\ 
	& 782 & 724 & 731 & 711 & 767 & 1648 & 1595 & 1562 & 1540 & 1783 & 1869 & 1658 & 748 \\ 
	& 787 & 729 & 732 & 710 & 771 & 1683 & 1572 & 1558 & 1547 & 1781 & 1874 & 1671 & 746 \\ 
	& 781 & 727 & 734 & 705 & 752 & 1676 & 1586 & 1564 & 1508 & 1772 & 1834 & 1636 & 743 \\ 
	& 732 & 671 & 673 & 662 & 724 & 1534 & 1459 & 1437 & 1449 & 1665 & 1759 & 1564 & 698 \\ 
	& 781 & 719 & 725 & 708 & 768 & 1656 & 1572 & 1547 & 1542 & 1776 & 1869 & 1661 & 744 \\ 
	& 781 & 727 & 735 & 704 & 754 & 1673 & 1589 & 1565 & 1506 & 1772 & 1837 & 1635 & 743 \\ 
	& 732 & 671 & 672 & 662 & 723 & 1534 & 1461 & 1435 & 1448 & 1667 & 1759 & 1563 & 698 \\ \hline
\end{tabular}
\vspace{0.5cm}
\caption{measured mesotectonic data. Notations: s: sign of the sample f: orientation of the fault r: rake of the stria}
\label{tab:ex_02}
\begin{tabular}{ r r r r r r r r r r r r r r r r r r r r}
\small
s &1  & 2 & 3 & 4 & 5 & 6 & 7 & 8 & 9 & 10 & 11 & 12 & 13 & 14 & 15 & 17 & 18 & 19\\\hline
f &155&150& 84& 96&356&354&350&160&180&  3 &340 & 162& 162& 350&  2 & 354& 357& 165\\
  & 75& 75& 80& 66& 72& 85& 75& 85& 85&  85& 82 &  80&  82&  76&  78&  88&  64&  80\\\hline
r	&16 & 5 &175&162& 20& 11&  5&171&170& 178&   5& 165& 178& 178&  10&  10&  20& 165\\
\end{tabular}
\end{table}



\end{document}